\documentclass{article}
\usepackage{spconf,amsmath,graphicx,svg}
\graphicspath{{./images/}}
\usepackage{algorithm}
\usepackage{algpseudocode}

\ninept
\title{speaker- and age-invariant training for child acoustic modelling using adversarial multi-task learning}
\name{Mostafa Shahin, Julien Epps, and Beena Ahmed}
\address{School of Electrical Engineering and Telecommunications, University of New South Wales, Sydney, Australia}
\begin{document}
%
\maketitle
\begin{abstract}
One of the major challenges in acoustic modelling of child speech is the rapid changes that occur in the children’s articulators as they grow up, their differing growth rates and the subsequent high variability in the same age group. These high acoustic variations along with the scarcity of child speech corpora have impeded the development of a reliable speech recognition system for children. In this paper, a speaker- and age-invariant training approach based on adversarial multi-task learning is proposed. The system consists of one generator shared network that learns to generate speaker- and age-invariant features connected to three discrimination networks, for phoneme, age, and speaker. The generator network is trained to minimize the phoneme-discrimination loss and maximize the speaker- and age-discrimination losses in an adversarial multi-task learning fashion. The generator network is a Time Delay Neural Network (TDNN) architecture while the three discriminators are feed-forward networks. The system was applied to the OGI speech corpora and achieved a 13
\end{abstract}
\section{Introduction}
\label{sec:intro}

Despite the enormous improvement in the acoustic modelling of adult speech over the last few decades, less progress has been made on the acoustic modelling of child speech. 

Automatic speech recognition systems trained on adult speech have shown a dramatic degradation in performance when tested on child speech due to linguistic and acoustic mismatches between adult and child speech \cite{Potamianos2003}. Children have higher fundamental and formant frequencies due to their smaller vocal cords and shorter vocal tract \cite{Mugitani2009}. Furthermore, the shape of the vocal tract changes rapidly as children grow up and their ability to correctly pronounce speech sounds improves. This leads to wider intra- and inter-speaker variations compared with adult speech \cite{Fitch1999,Lee1999}. 

Several approaches were initially proposed to handle variations in child speech at the feature level such as Vocal Tract Length Normalization (VTLN) \cite{Potamianos2003,Potamianos1997,Hagen2003,Serizel2014}, Stochastic Feature Mapping (SFM) \cite{Fainberg}, and Pitch Adaptive Mel-Frequency Cepstral Coefficient (PAMFCC) \cite{Shahnawazuddin2016}. Recently, deep learning techniques have become state-of-the-art in speech recognition systems, however, training such models needs a considerably large amount of speech data which is not available for child speech. Therefore, different domain adaptation and data augmentation techniques have been explored that incorporate both adult and child speech corpora such as teacher-student domain adaptation \cite{Li2017}, transfer learning \cite{Serizel,GurunathShivakumar2020}, and Multi-Task Learning (MTL) \cite{Wang2018}. 

Several works have studied the effect of age on the performance of child speech recognition. As expected, the performance degraded with a decrease of age when either an acoustic model trained on adult speech \cite{Potamianos1997} or age-specific acoustic model \cite{Yeung} was used. Due to the limited availability of child speech corpora, it is hard to train an accurate acoustic model for each age range. In \cite{Potamianos2003} acoustic model adaptation was used to adapt an adult acoustic model to the different age ranges. An age-dependent speaker normalisation technique was proposed in \cite{Guo} using subglottal resonances. 

Adversarial multi-task learning \cite{Ganin2015} has been used in literature for speaker invariant training of adult speech \cite{Saon2017,Meng}. However, acoustic variations in child speech are caused by both speaker and age variations due to the rapid, non-uniform growth of their articulators. In this paper, we thus investigated whether an adversarial multi-task learning approach can be used to alleviate the effect of both speaker and age acoustic variations in child speech. To achieve this, we proposed a system that uses two adversarial tasks, one for age and one for speaker discrimination, to generate speaker- and age-invariant features. Unlike most existing adversarial multi-task learning architectures where only one adversarial task was learnt jointly with the main task, the proposed architecture uses two adversarial tasks that were simultaneously trained along with the main phonetic discrimination task. Moreover, this is the first work to address both age and speaker variations in child speech using adversarial training. The system is validated using child speech corpus with a large number of speakers distributed over 11 age groups.

\section{Method}
\label{sec:meth}

Since Goodfellow presented the Generative Adversarial Network (GAN) as a novel method to generate samples from a target distribution \cite{goodfellow2020generative}, a variety of adversarial learning techniques have been proposed in literature including adversarial multi-task learning \cite{Ganin2015}. In traditional multi-task learning, a second different but relevant task is trained sharing part of the network, with the primary task to improve the generalization of the primary task \cite{Caruana1997}. In contrast, in adversarial multi-task learning, the shared network is learnt adversarially to the secondary task, i.e., to not discriminate between classes in the secondary task, resulting in representations invariant to the secondary task. Adversarial training has been successfully utilized to improve the robustness of speech recognition systems against noisy environments \cite{Shinohara2016}, speaker variations \cite{Saon2017,Meng}, and accent variation \cite{Sun2018}. Here we leverage adversarial multi-task learning to handle the high speaker- and age-variations in child speech.

\begin{figure}[h]
    \centering
    \includesvg[inkscapelatex=false, width=\columnwidth]{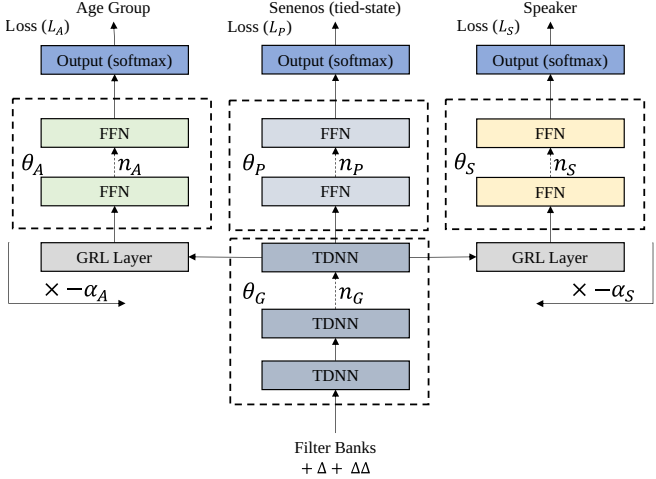}
    \caption{\textit{Adversarial multi-task learning architecture consisting of four networks: the generative network (G) of $\eta_G$ TDNN layers and trainable parameters $\theta_G$, the phoneme recognition network (P) of $\eta_P$ feed forward layers and trainable parameters $\theta_P$, the
speaker discrimination network (S) of $\eta_S$ feed forward layers and $\theta_S$ trainable parameters, and the age group discrimination network (A) of $\eta_A$ feed forward layers and $\theta_A$ trainable parameters.}}
    \label{fig:fig1}
\end{figure}

Figure \ref{fig:fig1} depicts our proposed architecture of the adversarial multi-task learning network. The architecture consists of four subnetworks: the generative network (G) which is the core network and acts as the feature extraction network, the phoneme recognition network (P) which is trained to classify senones, the speaker discrimination network (S) which is trained to discriminate between speech from different speakers, and the age group discrimination network (A) which is trained to discriminate between different age groups. The G network consists of $\eta_G$ Time Delay Neural Network (TDNN) layers similar to the one proposed in \cite{Snyder2016} with $\theta_G$ trainable parameters, while the networks P, S, and A are formed from $\eta_A$, $\eta_S$, and $\eta_A$ feed forward layers with $\theta_A$, $\theta_S$, and $\theta_A$ trainable parameters respectively. Moreover, the network has three output layers associated with the P, S, and A networks with output neurons equal to the number of tied-state triphones (senones), speakers, and age
groups respectively.

As shown in Figure \ref{fig:fig1}, the G network is shared among the three tasks, while each of the P, S, and A networks is associated with a separate task. Therefore, the cross-entropy loss function of each task is computed as follows

\begin{equation}
    L_P(\theta_G,\theta_P)=-\sum_i P(y_P^{(i)}|x^{(i)};\theta_G,\theta_P)
    \label{eq:eq1}
\end{equation}

\begin{equation}
    L_S(\theta_G,\theta_S)=-\sum_i P(y_S^{(i)}|x^{(i)};\theta_G,\theta_S)
    \label{eq:eq2}
\end{equation}

\begin{equation}
    L_A(\theta_G,\theta_A)=-\sum_i P(y_A^{(i)}|x^{(i)};\theta_G,\theta_A)
    \label{eq:eq3}
\end{equation}
where $x^{(i)}$ is the input feature vector of sample $i$, and $y_P^{(i)}$, $y_S^{(i)}$, and $y_A^{(i)}$ are the associated senones, speaker, and age group labels. During training, the P, S, and A parameters were updated such that their associated loss are minimized as follows

\begin{equation}
    \theta_P \gets \theta_P-\eta  \frac{\partial L_P}{\partial\theta_P}
    \label{eq:eq4}
\end{equation}

\begin{equation}
    \theta_S \gets \theta_S-\eta  \frac{\partial L_S}{\partial\theta_S}
    \label{eq:eq5}
\end{equation}

\begin{equation}
    \theta_A \gets \theta_A-\eta  \frac{\partial L_A}{\partial\theta_A}
    \label{eq:eq6}
\end{equation}

The G network is trained to minimize the senones classification loss and maximize both the speaker and age group discrimination losses allowing the network to generate features that are invariant to speaker and age variations. The parameters of the G network are therefore updated as follows

\begin{equation}
    \theta_G \gets \theta_G-\eta(\frac{\partial L_P}{\partial \theta_G}-\alpha_S\frac{\partial L_S}{\partial \theta_G}-\alpha_A\frac{\partial L_A}{\partial \theta_G})
    \label{eq:eq7}
\end{equation}
where $\eta$ is the learning rate and $\alpha_S$ and $\alpha_A$ are positive scalars denoting the portions of the speaker and age group adversarial tasks’ gradients backpropagated to update the generator’s parameters. This idea was achieved using the Gradient Reversal Layer (GRL) proposed by Ganin \cite{Ganin2015a}. As shown in Figure \ref{fig:fig1}, the GRL layer works as an identity-mapping in the forward path while in the backpropagation path it reverses and scales gradient by multiplying it with $-\alpha$

Equations (\ref{eq:eq5}), (\ref{eq:eq6}), and (\ref{eq:eq7}) demonstrate that the generator and the speaker and age discriminators play a minimax game. The generator tries to maximize the losses of the speaker and age classification tasks ($L_S$ and $L_A$) while the speaker and age discriminator networks try to minimize these losses. Most existing implementations of adversarial multi-task learning in speech acoustic modelling utilise only one adversarial task [24], In the implementation of our proposed architecture, two adversarial tasks were trained simultaneously along with the main task.

\begin{algorithm}
\caption{ADVERSARIAL MULTI-TASK LEARNING}
\label{alg:alg1}
\begin{algorithmic}
\Require Features: $X={x^{(1)},x^{(2)},\ldots,x^{(n)}}$
\Require Senon Label Sequence: $y_P={y_P^{(1)},y_P^{(2)},\ldots,y_P^{(n)}}$
\Require Speaker Label Sequence: $y_S={y_S^{(1)},y_S^{(2)},\ldots,y_S^{(n)}}$
\Require Age Label Sequence: $y_A={y_A^{(1)},y_A^{(2)},\ldots,y_A^{(n)}}$
\For {$r=0,1,\ldots,N-1$}
\For {$Epoch=1,2,\ldots,m$}
\State $L_P(\theta_G,\theta_P)=-\sum_i \log P(y_P^{(i)}|x^{(i)};\theta_G,\theta_P)$
\State \text{Compute} $\frac{\partial L_P}{\partial\theta_P}$ \text{\&} $\frac{\partial L_P}{\partial\theta_G}$
\State $\theta_P \gets \theta_P-\eta\frac{\partial L_P}{\partial \theta_P}$
\State $\theta_G \gets \theta_G-\eta\frac{\partial L_P}{\partial \theta_G}$
\EndFor
\For {$Epoch=1,2,\ldots,m$}
\State $L_S(\theta_G,\theta_S)=-\sum_i \log P(y_S^{(i)}|x^{(i)};\theta_G,\theta_S)$
\State $L_A(\theta_G,\theta_A)=-\sum_i \log P(y_A^{(i)}|x^{(i)};\theta_G,\theta_A)$
\State \text{Compute} $\frac{\partial L_S}{\partial\theta_S}$ \text{\&} $\frac{\partial L_A}{\partial\theta_A}$
\State $\theta_S \gets \theta_S-\eta\frac{\partial L_S}{\partial \theta_S}$
\State $\theta_A \gets \theta_A-\eta\frac{\partial L_A}{\partial \theta_A}$
\EndFor
\For {$Epoch=1,2,\ldots,m$}
\State $L_P(\theta_G,\theta_P)=-\sum_i \log P(y_P^{(i)}|x^{(i)};\theta_G,\theta_P)$
\State $L_S(\theta_G,\theta_S)=-\sum_i \log P(y_S^{(i)}|x^{(i)};\theta_G,\theta_S)$
\State $L_A(\theta_G,\theta_A)=-\sum_i \log P(y_A^{(i)}|x^{(i)};\theta_G,\theta_A)$
\State \text{Compute} $\frac{\partial L_P}{\partial\theta_P}$ \text{\&} $\frac{\partial L_S}{\partial\theta_S}$ \text{\&} $\frac{\partial L_A}{\partial\theta_A}$
\State $\theta_G \gets \theta_G-\eta(\frac{\partial L_P}{\partial \theta_G}-\alpha_S\frac{\partial L_S}{\partial \theta_G}-\alpha_A\frac{\partial L_A}{\partial \theta_G})$
\EndFor
\EndFor
\end{algorithmic}
\end{algorithm}

The training procedure is summarised in Algorithm \ref{alg:alg1}. Firstly, both generator (G) and senone classification (P) networks are trained for $m$ epochs to minimise the senone output loss $L_P$ allowing G to generate senone-discriminative features. In the next $m$ epochs, the
network G is frozen, so that its parameters are not updated, and both the speaker and age group discrimination networks trained to minimize their associated losses ($L_S$ and $L_A$). In the final $m$ epochs, the P, S, and A networks are frozen, and the generator is trained to minimize the senone classification loss $L_P$ and maximise the speaker and age discriminator losses ($L_S$ and $L_A$). Therefore, the generator learns to deceive the well-trained speaker and age discriminators resulting in speaker- and age-invariant features. These steps are typically repeated $N$ times.

\section{EXPERIMENTAL SETUP}
\label{sec:ExpSet}
\subsection{Dataset}
\label{ssec:data}

The Oregon Graduate Institute (OGI) kids’ speech corpus \cite{Shobaki2000} was used in this work. It was collected from 1110 children distributed over 11 age groups based on their school grade from kindergarten to grad 10. Most of the recordings are scripted, i.e., each child read a set of prompt words and short sentences. Additionally, spontaneous speech was elicited by answering a set of open questions. In this work only the scripted part of the dataset was utilized. The speech corpus was split into three parts, 794 speakers for training, 158
speakers for developing, and 158 speakers for testing.

\subsection{Training procedure}
\label{ssec:train}

Each speech sample was divided into 25 msec frames with 15 msec overlap. Each frame was first multiplied by a Hamming window and 40 Mel-scale cepstral coefficients were extracted from each frame. 

The tied-state triphones (senones) alignment was obtained
using a GMM-HMM initial acoustic model trained on the same
speech corpus. Therefore, three labels were assigned to each frame, the speaker index, the age group index, and the senone index. The baseline model consisted of 5 TDNN layers with delays of {-2,-1,0,1,2}, {-1,2}, {-3,3}, {-3,3}, and {-7,2} each with 1024 neurons followed by a feed forward fully connected layer of 1024 neurons. The effective temporal context of the model is 39 frames before and after the underlying frame (-39, +39). The final output layer had 1360 neurons representing the number of senones.

For the adversarial multi-task learning architecture depicted in Figure \ref{fig:fig1}, the G network was similar to the baseline model with the same 5 TDNN layers while each of the P, S, and A networks had one feed forward fully connected layer in addition to the last output layer. The numbers of neurons of the output layers of the P, S, and A discriminators were 1360, 794, and 11 representing the number of senones, training speakers, and age groups respectively.

As shown in Figure 1, the P discriminator receives input directly from the output of the last TDNN layer of the G network. On the contrary, both S and A discriminators receive inputs through GRL layers with scaling factors $\alpha_s$ and $\alpha_A$ respectively. The scaling factors $\alpha_s$ and $\alpha_A$ were gradually increased from 0 to 0.01 as follows

\begin{equation}
    \alpha_S=\alpha_A=\frac{r}{N-1}\times \alpha \;\;\;\; r=0,1,\ldots,N-1
    \label{eq:eq8}
\end{equation}
where $r$,$N$  are the repeat index and the total number of repeats respectively as explained in Algorithm \ref{alg:alg1} and $\alpha$ is the maximum value that is reached at repeat $N-1$.

The effect of each adversarial task separately was further investigated by training two separate models with each having only one adversarial task. The acoustic model was evaluated in terms of the Word Error Rate (WER) in an ASR task. The language model used was a 4-gram model trained on the transcripts of the OGI data.

\section{Results}
\label{sec:res}
Four different models were trained: the baseline model, the AGE model, the SPK model and the AGE+SPK model.

In the AGE model, the speaker discriminator (S) was removed, and the G network was trained adversarially against the age group discriminator (A) only. In contrast, for the SPK model, only the speaker discriminator was used in the adversarial update of the G network parameters. Finally, the AGE+SPK model utilized both discriminators. All four acoustic models were used to achieve ASR and were tested using the test and development subsets of the OGI dataset. For all models $\alpha_{max}$ was set to 0.01.

\begin{figure}[h]
    \centering
    \includesvg[inkscapelatex=false, width=\columnwidth]{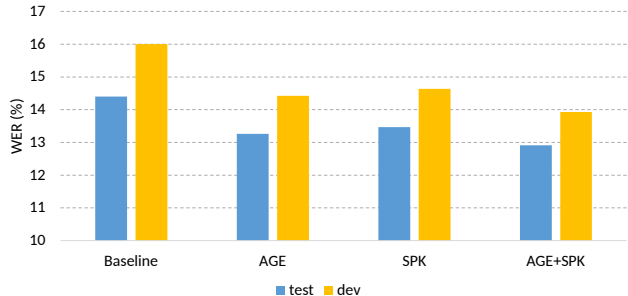}
    \caption{\textit{The effect of the adversarial multi-task learning on the WER of the ASR trained on the OGI speech corpus. The baseline model was trained with no adversarial update to the parameters. The AGE and SPK with only the age discriminator or the speaker discriminator were utilized respectively, whereas AGE+SPK when both discriminators were used.}}
    \label{fig:fig2}
\end{figure}

Figure \ref{fig:fig2} shows the WER of the test and development subsets of the four models. The baseline model achieved WERs of 14.4\% and 16\% respectively. Training the model adversarially against the speaker discriminator (SPK) reduced the WER by ~8\% and ~10\% to 13.47\% and 14.64\% for the test and development subsets respectively. The AGE model gave slightly better improvement compared with the SPK model, with WERs of 13.26\% and 14.42\% for the test and development subsets respectively. The best performance was obtained when both speaker and age discriminators were used with WER relative reductions of ~10\% and ~13\% for the test and development subsets respectively. These results indicate that the age variation has more influence on the performance of the acoustic model than the speaker variation.

\begin{figure*}[ht]
    \centering
    \includesvg[inkscapelatex=false, width=\textwidth]{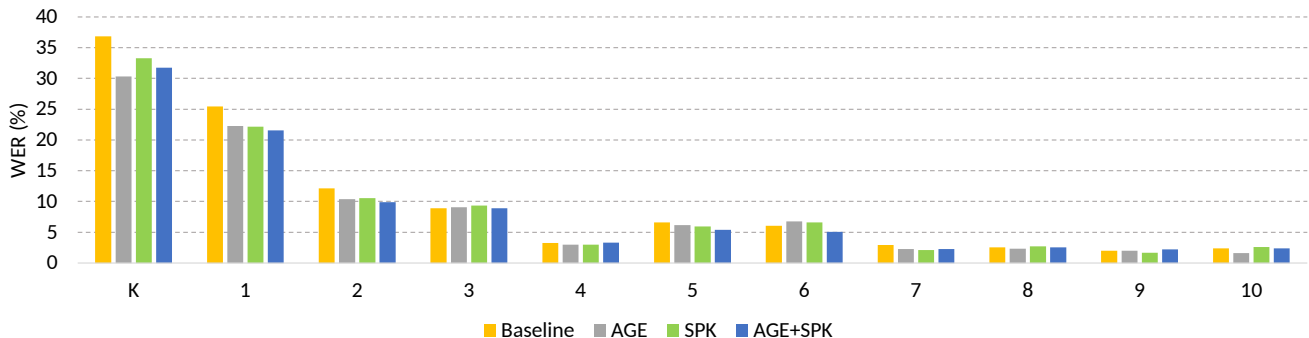}
    \caption{\textit{Age breakdown of the performance of the four models, Baseline, AGE, SPK, and AGE+SPK}}
    \label{fig:fig3}
\end{figure*}

To better interpret these results, we broke them down to age
level as depicted in Figure \ref{fig:fig3}. It can be seen that WER reduced
significantly in grades K and 1 compared to elder ages. It is noted also that at grade K using only AGE adversarial training achieved the lower WER than using SPK or AGE+SPK models. This
indicates that there is a high degree of discrepancy between children at kindergarten and elder ages caused mainly due to the age variations which have been successfully alleviated using AGE adversarial training.

\begin{figure}[h]
    \centering
    \includesvg[inkscapelatex=false, width=\columnwidth]{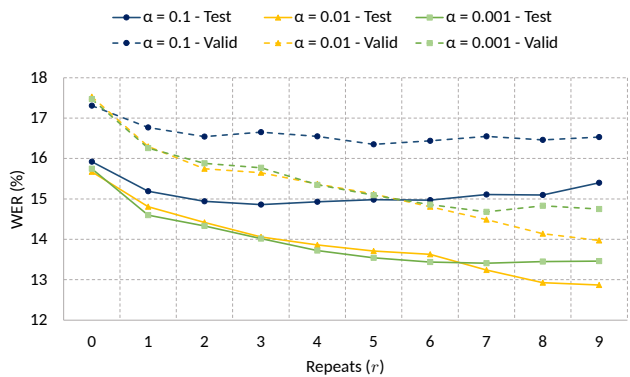}
    \caption{\textit{The WER of the AGE+SPK model at different values of $\alpha$ parameter over each repeat.}}
    \label{fig:fig4}
\end{figure}

Figure 4 shows the effect of the $\alpha$ parameter which represents the portion of the adversarial gradient backpropagated to update the generator network parameters. Three values of $\alpha$ were tested, namely 0.1, 0.01, and 0.001. Figure \ref{fig:fig4} also demonstrates the convergence of the model with the repeats (see Algorithm \ref{alg:alg1}). The WER of both the validation and testing sets are computed after each
repeat. Note that as formulated in equation (\ref{eq:eq8}), at repeat 0 the values of $\alpha_S$ and $\alpha_A$ equals to 0 for all $\alpha$ values and then increased gradually at each repeat, therefore, the WER of the three $\alpha$ values starts from the same point at ~15.5\% and 17.5\% WER for the testing and validation sets respectively.

The results show that when using $\alpha=0.1$, the WER slightly decreased at first repeat when the effective values of $\alpha_S$ and $\alpha_A$ were ~0.01 and then almost saturated as the effective values of $\alpha_S$ and $\alpha_A$ increased. On the other hand, at $\alpha=0.001$ and 0.01, the WER decreased at each repeat and almost saturated at repeat 5 with ~13.5\% WER (test set) and ~15\% WER (validation), and at repeat 8 with ~13\% WER (test set), and ~14\% WER (validation), for $\alpha=0.001$ and $\alpha=0.01$ respectively. These results demonstrate that selecting the amount of reversed gradient used to update the feature generator network is crucial in the effectiveness of the training.

\section{CONCLUSION}
\label{sec:con}

In this paper, a speaker- and age-invariant training approach was proposed leveraging the adversarial multi-task learning paradigm for child acoustic modelling to generate the speaker and age acoustic variations. A feature extraction network was learnt to generate phonetic discriminative features that are invariant to speaker and age variations. This was achieved by training the generator to minimize the phoneme discrimination loss and, at the same time, maximize the speaker and age discrimination losses. This approach boosted the ASR word recognition accuracy from ~85\% to ~87\% when tested on the Oregon Graduate Institute (OGI) test set.

This shows that the two-way adversarial multi-task learning was effective in combining data from different ages for the training of age-independent acoustic model. The age analysis showed that the improvement was significant in recognizing speech of kindergarten children compared to elder ages.

\bibliographystyle{IEEEbib}
\bibliography{ICASSP_2023}

\end{document}